\documentclass[prd,twocolumn,superscriptaddress,showpacs,preprintnumbers,amssymb,floatfix]{revtex4-2}

\usepackage[utf8]{inputenc}
\usepackage[T1]{fontenc}
\usepackage{booktabs}
\usepackage{mathrsfs}
\usepackage{graphicx}% Include figure files
\usepackage{dcolumn}% Align table columns on decimal point
\usepackage{bm}% bold math
\usepackage{lipsum}
\usepackage{graphics}
\usepackage{bbm}
\usepackage{color}
\usepackage{microtype}
\usepackage{IEEEtrantools}
\usepackage{braket}
\usepackage{mathrsfs}
\usepackage{amsthm}
\usepackage{enumitem}
\usepackage[normalem]{ulem}
\usepackage[pdftex,breaklinks,colorlinks,
linkcolor=blue,
citecolor=teal,
anchorcolor=red,
urlcolor=cyan]{hyperref}
\usepackage{orcidlink}
\usepackage{mathtools}
\usepackage{fontawesome}
\graphicspath{{Figures/}}

\begin{document}

\title{Operator-Level Quantum-Classical Correspondence in Relativistic Quantum Theory and Curved Spacetime}

\author{Pankaj Sheoran\orcidlink{0000-0001-8283-8744}
}
\email{pankaj.sheoran@vit.ac.in}
\affiliation{Department of Physics, School of Advanced Sciences, Vellore Institute of Technology, Tiruvalam Rd, Katpadi, Vellore, Tamil Nadu 632014, India}

\author{Gopal Kashyap\orcidlink{0000-0001-9001-4905}
}
\email{gopal.hari@vit.ac.in}
\affiliation{Department of Physics, School of Advanced Sciences, Vellore Institute of Technology, Tiruvalam Rd, Katpadi, Vellore, Tamil Nadu 632014, India}

\author{Sanjay Siwach\orcidlink{0000-0001-8881-1315}
}
\email{sksiwach@hotmail.com}
\affiliation{Department of Physics, Institute of Science, Banaras Hindu University, Varanasi-221005, India}

\date{\today}

\begin{abstract}

  In a recent work \cite{Shaikh:2026nsi}
% [arXiv:2601.10423]
, it was shown that for a non-relativistic particle, the operator equations of motion can be cast in the same differential form as Newton’s equation of motion. This establishes a correspondence between quantum operator dynamics and classical mechanics at the level of the equations of motion. In this work, we investigate whether this correspondence extends to quantum field theory and linearized gravity.
We explicitly show that the Heisenberg equations of motion for quantum field operators reproduce the corresponding classical equations of motion for a broad class of field theories. The Klein-Gordon equation, Maxwell's equations, Yang-Mills equations, and linearized Einstein equations are shown to arise at the operator level without taking the limit $\hbar\rightarrow0$. We then extend the analysis to curved spacetime and show that the quantum field operators satisfy the corresponding covariant field equations at the operator level. In the weak-field approximation, we show that the Heisenberg evolution of the metric-perturbations reproduces the linearized Einstein equations at the operator level. For a quantum particle in curved spacetime, the Heisenberg equations for position and momentum reproduce classical geodesic motion in the semiclassical limit, thereby extending the operator-level quantum-classical correspondence from fields to particle dynamics. 

 \end{abstract}

\maketitle

\section{Introduction}

Quantum mechanics is one of the most successful physical theories in the modern era and provides an accurate account of the microscopic phenomena. From the hydrogen atom to collisions at high energies, all of its predictions have been found very precise \cite{Peskin:1995,Weinberg:1995,Itzykson:1980}. Even after its great success for the last hundred years, the connection between quantum mechanics and classical mechanics has always been difficult. Classical mechanics describes the motions of macroscopic bodies and can be taken as an approximation of quantum mechanics under specific circumstances \cite{Ehrenfest:1927,holland1993quantum}. But, how precisely this approximation is obtained through the theory of quantum mechanics is still an open question. Niels Bohr's correspondence principle at the beginning of the twentieth century provides a clue in this direction. According to this principle, quantum mechanics leads to classical mechanics in the limit of large quantum numbers \cite{Hassoun:1989}. In semi-classical approach using the Wentzel-Kramers-Brillouin approximation, the Planck’s constant is considered as a small parameter, and one recovers the classical equations as the leading-order approximation \cite{Greensite:2017}. In the Feynman path integral approach, the classical principle of minimum action appears in the limit $\hbar \to 0$ where the fast oscillations of the phase $e^{iS/\hbar}$ make all non-stationary paths zero \cite{Feynman:1965}. More extreme ideas include the possibility that nonlinear changes should be made to quantum mechanics in order to completely recover classical mechanics \cite{Bassi:2012bg, Penrose:1996cv}. An alternative position is presented by decoherence theory, which explains the process of appearance of classicality through interaction of quantum states with the environment \cite{Giulini:1996nw}. It has also been argued that quantum mechanics is an emergent theory \cite{Adler:2004}.

In the midst of such a wide-ranging set of possibilities, one recent insight has come into more precise focus: the time evolution equations for quantum observables in the Heisenberg picture can, under surprisingly general circumstances, be expressed in a way that is mathematically equivalent to the equations of classical mechanics \cite{Shaikh:2026nsi}.
%Shaikh and Qureshi have shown this in their work \cite{Shaikh:2026nsi} for a single non-relativistic particle subjected to an analytic potential. 
Starting from the Heisenberg equations for the position and momentum operators, they
obtained 
\begin{equation}
m\frac{d^{2}\hat{\mathbf{x}}(t)}{dt^{2}}=- V'\bigl(\hat{\mathbf{x}}(t)\bigr),
\end{equation}
which has the same form as Newton's second law, with the classical position and momentum replaced by their Heisenberg picture operator counterparts. This derivation did not use any approximations, no assumption of smallness of $\hbar$, nor limitation to specific quantum states.
The same considerations have been applied to charged particles in electromagnetic fields, resulting in the operator formulation of the Lorentz force, and to rotating coordinate systems, where Coriolis and centrifugal forces are deduced from the Heisenberg equations in the same way as in classical mechanics. Generalized derivation based on Hamilton’s equations has demonstrated that for any Hamiltonian, the Heisenberg equations simplify to
\begin{equation}
\frac{d\hat{q}_j}{dt} = \frac{\partial \hat{H}}{\partial \hat{p}_j}, \qquad \frac{d\hat{p}_j}{dt} = -\frac{\partial \hat{H}}{\partial \hat{q}_j},
\end{equation}
which have the same {differential} form as Hamilton's equations of classical mechanics, with the canonical variables replaced by the operators.

 It was noticed that no explicit $\hbar$ remains in the resulting equations of motion, and the dynamical evolution of quantum observables, governed by the Heisenberg equations, is formally identical to classical dynamics. What makes the two approaches different is not the form of the equations of observables but the fact that quantum observables do not commute and that the quantum state determines the expectation value.

Although this result may seem surprising both in terms of its generality and simplicity, it is explored in the context of non-relativistic quantum mechanics only. It is natural to wonder whether an analogous correspondence also holds in relativistic quantum field theory. Quantum field theory provides the framework for describing elementary particles and the electromagnetic, weak, and strong interactions \cite{Peskin:1995,Weinberg:1995,Itzykson:1980}. It incorporates special relativity, particle creation and annihilation, and the infinitely many degrees of freedom associated with fields. If the aforementioned correspondence is indeed an intrinsic property of nature, it should manifest itself also in this more universal framework. 
The problem is whether one can prove that the Heisenberg equations for quantum fields give rise to the corresponding classical equations of motion for these fields directly at the operator level, and to determine under which circumstances the explicit $\hbar$ dependence cancels without using any approximation.

 The emergence of classical dynamics from quantum field theory has been widely studied using coherent-state constructions
\cite{Lasha_2021,Lasha_2024} and functional methods or the semiclassical WKB approximation. Under suitable conditions, coherent-state expectation values can follow the classical field evolution, with deviations arising from the quantum corrections. The approach we take here is distinct from this line of thought. Rather than concentrating on the expectation values of observables or certain states, we examine the operator equations themselves. 
The central question we want to answer is whether the Heisenberg field operators satisfy equations which reproduce the corresponding classical field equations at the operator level. Establishing such a result would extend the operator-level correspondence found in nonrelativistic single-particle systems to relativistic fields and interactions.

In the context of quantum gravity, which remains in search of a consistent theory, understanding the relation between quantum dynamics and underlying spacetime geometry is of crucial importance \cite{Birrell:1982,Wald:1994,Mukhanov:2007}. Within the weak-field approximation, if the Heisenberg evolution of the metric perturbations reproduce the linearized Einstein equations, it would show that quantum and classical gravity share the same dynamical structure at this level. The quantum character would still reside in the noncommuting operators and the quantum states. 

In this work, we develop this connection more systematically. We first present the non-relativistic case in the framework of Dirac notation to establish the conceptual groundwork. We then consider scalar and Dirac fields in flat spacetime and derive the corresponding Klein-Gordon and Dirac equation using the Heisenberg evolution of the operators. For gauge theories, we obtain the operator form of {Dirac's} equations in quantum electrodynamics (QED) and equations for a non-Abelian gauge field coupled to Dirac matter. The analysis is further extended to scalar, Dirac, and electromagnetic fields in curved spacetime. We also consider semiclassical gravity and weak-field quantum gravity, showing that the Heisenberg evolution of the metric perturbation yields the linearized Einstein equations at the operator level. Finally, for a quantum particle in curved spacetime, we derive the exact first-order Heisenberg equations for position and momentum and show that their semiclassical limit reproduces the geodesic equation. Within the stated assumptions, the field equations follow directly from the corresponding Hamiltonians and canonical (anti)commutation relations, without relying on a classical-state approximation.
Together, these results extend the operator-level correspondence identified for nonrelativistic single-particle systems in ~\cite{Shaikh:2026nsi} to relativistic quantum fields and complement the relativistic-particle analysis of ~\cite{Kong2024}, by
treating relativistic quantum fields and interactions in flat and curved spacetime, including the Abelian and non-Abelian gauge theories and the weak-field gravity.

The structure of the paper is as follows. Section \ref{sec:HP_n_S_field} briefly {reviews the Heisenberg picture and examines the dynamics of free and interacting scalar fields} in flat spacetime. Section \ref{sec:D_field} discusses the Dirac field and is divided into two subsections: one devoted to the free Dirac field and the other to the interacting Dirac field, using the Yukawa interaction as an example. In Sec. \ref{sec:Gauge_field}, we analyze flat-spacetime gauge field theories, with subsections dedicated to quantum electrodynamics and non-Abelian Yang-Mills theory. We extend the analysis to quantum fields in curved spacetime in Sec. \ref{sec:qf_curved_st}, considering scalar, Dirac, and electromagnetic fields, as well as gravity coupled to matter fields. We then study the implications in the context of linearized gravity in Sec. \ref{Lin_Ein}. We also show that the motion of a quantum particle in curved spacetime, starting from the Heisenberg equation of motion, leads explicitly to the classical geodesic motion in curved spacetime. Finally, Sec. \ref{sec:conclusions} concludes the paper by summarizing our key observations and results.
Throughout this work, we adopt the metric signature $\eta_{\mu\nu}=\text{diag}(+1,-1,-1,-1)$. We set $c=1$ for simplicity while retaining $\hbar$ explicitly.

\section{Heisenberg Picture and Scalar Field Dynamics in Flat Spacetime}\label{sec:HP_n_S_field}

To begin with, in this section, we briefly review the 
Schr\"odinger and Heisenberg pictures in the context of quantum field theory (QFT). We then apply the Heisenberg formalism for QFT to a real scalar field, and derive the Heisenberg equations of motion for the field operator and its conjugate momentum, and finally obtaining the Klein–Gordon equation (and its interacting generalization) at the operator level.
It is well known that in the Schr\"odinger picture, the state vectors carry the full time dependence, while the operators are time-independent (unless they have explicit time dependence from external sources). The time evolution of a state vector $\ket{\psi_S(t)}$ is governed by the Schr\"odinger equation as,
\begin{equation}\label{eq:Sch_eq}
i\hbar \frac{d}{dt} \ket{\psi_S(t)} = \hat{H}_S \ket{\psi_S(t)},
\end{equation}
where $\hat{H}_S$ is the time-independent Schr\"odinger-picture Hamiltonian. The formal solution of the above Eq. (\ref{eq:Sch_eq}) is given by,
\begin{equation}
\ket{\psi_S(t)} = e^{-i\hat{H}_S t/\hbar} \ket{\psi_S(0)}.
\end{equation}

Schr\"odinger-picture field operators $\hat{\phi}_S(\mathbf{x})$ are defined at a fixed initial time (usually $t=0$) and hence, do not evolve in time. They satisfy the following canonical commutation relation,
\begin{equation}
[\hat{\phi}_S(\mathbf{x}), \hat{\pi}_S(\mathbf{y})] = i\hbar \delta^{(3)}(\mathbf{x} - \mathbf{y}),
\end{equation}
where $\hat{\pi}_S(\mathbf{x})$ is the conjugate momentum operator.

On the other hand, the Heisenberg picture is defined using a unitary transformation that transfers the time dependence from the state vectors to the operators. The Heisenberg picture state vector is defined as,
\begin{equation}
\ket{\psi_H} = e^{i\hat{H}_S t/\hbar} \ket{\psi_S(t)} = \ket{\psi_S(0)}.
\end{equation}
Thus, in the Heisenberg picture, the state vectors are time-independent. The Heisenberg picture field operator is defined as,
\begin{equation}\label{eq:Heisen_field_O}
\hat{\phi}_H(t,\mathbf{x}) = e^{i\hat{H}_S t/\hbar} \hat{\phi}_S(\mathbf{x}) e^{-i\hat{H}_S t/\hbar}.
\end{equation}
It is clearly seen from the above equation that at $t=0$, the Heisenberg and Schr\"odinger picture operators coincide.
The time evolution equation for the Heisenberg picture field operator $\hat{\phi}_H(t,\mathbf{x})$ reads as,
\begin{equation}\label{eq:Heisen_eq}
\frac{\partial \hat{\phi}_H}{\partial t} = \frac{i}{\hbar} [\hat{H}_H, \hat{\phi}_H(t,\mathbf{x})].
\end{equation}

This is the fundamental equation governing the time evolution of Heisenberg picture field operators.
The equal-time commutation relations in the Heisenberg picture follow from the Schr\"odinger picture relations and are given by
\begin{align}
[\hat{\phi}_H(t,\mathbf{x}), \hat{\pi}_H(t,\mathbf{y})]=[\hat{\phi}(t,\mathbf{x}), \hat{\pi}(t,\mathbf{y})] = i\hbar \delta^{(3)}(\mathbf{x} - \mathbf{y}).
\label{commutator}
\end{align}
where we have dropped the $H$ subscript in the first equality. Henceforth, we shall drop the subscript $H$ for all the operators in the Heisenberg picture.   

\subsection{Non-interacting scalar field} 

Let us consider the Lagrangian density of a massive real scalar field in flat spacetime, given by
\begin{equation}\label{eq:action_flat}
\mathcal{L}
= 
\frac{1}{2}\,\partial_\mu\phi\,\partial^\mu\phi
- \frac{1}{2}\frac{m^2}{\hbar^2}\phi^2.
\end{equation}
The conjugate momentum is obtained as
\begin{equation}\label{eq:conj_mom_flat}
\pi = \frac{\partial\mathcal{L}}{\partial\dot{\phi}} = \dot{\phi},
\end{equation}
where the dot denotes the time derivative.
The Hamiltonian density is obtained by a Legendre transformation and is given as
\begin{equation}\label{eq:H_density_flat}
\mathcal{H} = \pi\dot{\phi} - \mathcal{L}
= \frac{1}{2}\pi^2 + \frac{1}{2}(\nabla\phi)^2
+ \frac{1}{2}\frac{m^2}{\hbar^2}\phi^2 .
\end{equation}

We now promote $\phi$ and $\pi$ to operators satisfying the equal-time commutation relation (\ref{commutator}). 
The operator form of the Hamiltonian of the system (spatial integral of Hamiltonian density) is given as
\begin{equation}\label{eq:H_scalar}
\hat{H} = 
\int d^3x \left( \frac{1}{2} \hat{\pi}^2 + \frac{1}{2} (\nabla \hat{\phi})^2 + \frac{1}{2} \frac{m^2}{\hbar^2} \hat{\phi}^2 \right).
\end{equation}
Now, using Eq. (\ref{eq:H_scalar}), the commutation relation on the right hand side of the Eq. (\ref{eq:Heisen_eq}) reads as 
\begin{align}\label{eq:H_P_com}
[\hat{H}, \hat{\phi}(t,\mathbf{x})] &= \int d^3y \left( \frac{1}{2} [\hat{\pi}^2(t,\mathbf{y}), \hat{\phi}(t,\mathbf{x})]\right. \nonumber\\
&+ \frac{1}{2} [(\nabla \hat{\phi}(t,\mathbf{y}))^2, \hat{\phi}(t,\mathbf{x})]\nonumber\\
&\left.+ \frac{1}{2} \frac{m^2}{\hbar^2} [\hat{\phi}^2(t,\mathbf{y}), \hat{\phi}(t,\mathbf{x})] \right).
\end{align}
Since $\hat{\phi}$ commutes with itself and with its spatial derivatives, only the $\hat{\pi}^2$ term in the Hamiltonian contributes.
Thus,
\begin{equation}
[\hat{H}, \hat{\phi}(t,\mathbf{x})] = \frac{1}{2} \int d^3y \, [\hat{\pi}^2(t,\mathbf{y}), \hat{\phi}(t,\mathbf{x})].
\end{equation}
Now, using the identity $[\hat{A}\hat{B}, \hat{C}] = \hat{A}[\hat{B}, \hat{C}] + [\hat{A}, \hat{C}]\hat{B}$ and the equal-time canonical commutation relation, we get
\begin{align}
[\hat{\pi}^2(t,\mathbf{y}), \hat{\phi}(t,\mathbf{x})] &= -2i\hbar \hat{\pi}(t,\mathbf{y}) \delta^{(3)}(\mathbf{y} - \mathbf{x}).
\end{align}
Integrating this over $\mathbf{y}$ implies,
\begin{align}
[\hat{H}, \hat{\phi}(t,\mathbf{x})] &= \frac{1}{2} \int d^3y \left( -2i\hbar \hat{\pi}(t,\mathbf{y}) \delta^{(3)}(\mathbf{y} - \mathbf{x}) \right) \nonumber \\
&= -i\hbar \hat{\pi}(t,\mathbf{x}).
\end{align}
Substituting into the Heisenberg Eq. (\ref{eq:Heisen_eq}), gives the operator relation,
\begin{equation} {\label{eq:Heisenberg1}}
\hat{\pi}(t,\mathbf{x}) = \frac{\partial \hat{\phi}(t,\mathbf{x})}{\partial t}.
\end{equation}
This is the quantum analog of the classical relation in Eq. (\ref{eq:conj_mom_flat}),
and correctly reflecting that $\hat{\phi}$ depends on time through its Heisenberg evolution.

The Heisenberg equation for conjugate momentum $\hat{\pi}$, is
\begin{equation}\label{eq:conj_p}
\frac{\partial \hat{\pi}(t,\mathbf{x})}{\partial t} = \frac{i}{\hbar} [\hat{H}, \hat{\pi}(t,\mathbf{x})].
\end{equation}
Now, similar to Eq. (\ref{eq:H_P_com}), we can write the commutation relation in the above equation as
\begin{align}\label{eq:H_p_com1}
[\hat{H}, \hat{\pi}(t,\mathbf{x})] &= \int d^3y \left( \frac{1}{2} [(\nabla \hat{\phi}(t,\mathbf{y}))^2, \hat{\pi}(t,\mathbf{x})]\right.\nonumber\\
&\left.+ \frac{1}{2} \frac{m^2}{\hbar^2} [\hat{\phi}^2(t,\mathbf{y}), \hat{\pi}(t,\mathbf{x})] \right).
\end{align}
Since $\hat{\pi}$ commute with itself, the $\hat{\pi}^{\,2}$ term does not contribute; only the gradient and mass terms remain. The first commutation relation in the above equation gives
\begin{align}
[(\nabla \hat{\phi}(t,\mathbf{y}))^2, \hat{\pi}(t,\mathbf{x})] &= -2i\hbar \nabla^2 \hat{\phi}(t,\mathbf{y})\delta^{(3)}(\mathbf{y} - \mathbf{x}),
\end{align}
and the second commutation relation gives
\begin{align}
[\hat{\phi}^2(t,\mathbf{y}), \hat{\pi}(t,\mathbf{x})] &= 2i\hbar \hat{\phi}(t,\mathbf{y}) \delta^{(3)}(\mathbf{y} - \mathbf{x}),
\end{align}
Using these two relations, Eq. (\ref{eq:H_p_com1}) becomes
\begin{align}
[\hat{H}, \hat{\pi}(t,\mathbf{x})] &= 
 -i\hbar \nabla^2 \hat{\phi}(t,\mathbf{x}) + i\hbar \frac{m^2}{\hbar^2} \hat{\phi}(t,\mathbf{x}).
\end{align}
Substituting into the Heisenberg equation for $\hat{\pi}$, we get
\begin{align}
\frac{\partial \hat{\pi}(t,\mathbf{x})}{\partial t} 
&= \nabla^2 \hat{\phi}(t,\mathbf{x}) - \frac{m^2}{\hbar^2} \hat{\phi}(t,\mathbf{x}).
\end{align}
Now, using the relation (\ref{eq:Heisenberg1}), we can write,  
\begin{equation}
\frac{\partial}{\partial t} \left( \frac{\partial \hat{\phi}}{\partial t} \right) = \nabla^2 \hat{\phi} - \frac{m^2}{\hbar^2} \hat{\phi},
\end{equation}
or
\begin{equation}
\left(\Box+\frac{m^{2}}{\hbar^{2}}\right)\hat{\phi}=0,
\end{equation}
where $\Box=\partial_{\mu}\partial^{\mu} =\partial_t^2-\nabla^2 $. This is the operator form of the Klein-Gordan equation. It establishes the quantum-classical correspondence by showing that the quantum field operator obeys an equation identical to its classical counterpart.

{The $\hbar$ introduced by the canonical commutation relation cancels with that in the Heisenberg equation. For a free massless scalar field, the equation reduces to
\begin{equation}
\Box\hat{\phi}=0,
\end{equation}
and no explicit $\hbar$ dependence remains in the operator equation of motion.
}
 
\subsection{Interacting scalar field} 

We now consider a real scalar field with a quartic self-interaction and the Lagrangian density is given by,
\begin{equation}\label{eq:action_flat_int}
\mathcal{L}
= 
\frac{1}{2}\,\partial_\mu\phi\,\partial^\mu\phi
- \frac{1}{2}\frac{m^2}{\hbar^2}\phi^2
- \frac{\lambda}{4 \hbar}\phi^4,
\end{equation}
where $\lambda$ is the quartic self-interaction coupling constant. The corresponding conjugate momentum is obtained as
\begin{equation}\label{eq:conj_mom_flat_int}
\pi = \frac{\partial\mathcal{L}}{\partial\dot{\phi}} = \dot{\phi},
\end{equation}
which is unchanged from the non-interacting case because the interaction term
does not contain $\dot{\phi}$. The Legendre transformation then gives
\begin{equation}\label{eq:H_density_flat_int}
\mathcal{H} 
= \frac{1}{2}\pi^2 + \frac{1}{2}(\nabla\phi)^2
+ \frac{1}{2}\frac{m^2}{\hbar^2}\phi^2
+ \frac{\lambda}{4\hbar}\phi^4 .
\end{equation}
Similar, to non-interacting case, we now promote $\phi$ and $\pi$ to operators satisfying the equal-time
commutation relations. The operator form of Hamiltonian of
interacting scalar field can be written as
\begin{equation}\label{eq:H_density_flat_int_op}
\hat H = \int d^3x\left(\frac{1}{2} \hat{\pi}^2 + \frac{1}{2} (\nabla \hat{\phi})^2
+ \frac{1}{2} \frac{m^2}{\hbar^2} \hat{\phi}^2
+ \frac{\lambda}{4 \hbar} \hat{\phi}^4\right).
\end{equation}

The Heisenberg Eq.~(\ref{eq:Heisenberg1}) for $\hat{\phi}$ remains unchanged because $\hat{\phi}^4$ commutes with $\hat{\phi}$. Thus, we still have $\partial \hat{\phi}/\partial t = \hat{\pi}$. For $\hat{\pi}$, we need an additional commutator,
\begin{align}
[\hat{\phi}^4(t,\mathbf{y}), \hat{\pi}(t,\mathbf{x})] &= 4\hat{\phi}^3(t,\mathbf{y}) [\hat{\phi}(t,\mathbf{y}), \hat{\pi}(t,\mathbf{x})],\nonumber\\
&= 4i\hbar \hat{\phi}^3(t,\mathbf{y}) \delta^{(3)}(\mathbf{y} - \mathbf{x}).
\end{align}
Therefore,
\begin{align}
[\hat{H}, \hat{\pi}(t,\mathbf{x})] &= -i\hbar \nabla^2 \hat{\phi}(t,\mathbf{x}) + i\hbar \frac{m^2}{\hbar^2} \hat{\phi}(t,\mathbf{x}) + \frac{\lambda}{ \hbar} \cdot i\hbar \hat{\phi}^3(t,\mathbf{x}) \nonumber \\
&= -i\hbar \nabla^2 \hat{\phi} + i\hbar \frac{m^2}{\hbar^2} \hat{\phi} + i \lambda \hat{\phi}^3.
\end{align}
The Heisenberg equation for $\hat{\pi}$ then gives
\begin{equation}
\frac{\partial \hat{\pi}}{\partial t} = \nabla^2 \hat{\phi} - \frac{m^2}{\hbar^2} \hat{\phi} - \frac{\lambda}{\hbar} \hat{\phi}^3.
\end{equation}

Using $\hat{\pi} = \partial \hat{\phi}/\partial t $, we obtain the nonlinear Klein-Gordon equation
\begin{equation}
\Box\hat{\phi}
+\frac{m^{2}}{\hbar^{2}}\hat{\phi}
+\frac{\lambda}{\hbar}\hat{\phi}^{3}=0.
\end{equation}
Thus, even in the massless limit,
\begin{equation}
\Box\hat{\phi}
+\frac{\lambda}{\hbar}\hat{\phi}^{3}=0,
\end{equation}
explicit $\hbar$ dependence remains through the interaction term in the present field normalization {implying that the interacting field has a non-trivial quantum dynamics}.

\section{Dirac Field Theory in Flat Spacetime}\label{sec:D_field}
In this section, we consider the Dirac field, which describes spin-$1/2$ fermions (e.g. electrons and quarks). 
We discuss the Dirac field, considering the two cases: $(i)$ free fields and $(ii)$ Dirac field in the presence of interactions.

\subsection{Free Dirac Field} \label{sec:free_D_field}

We begin with the Dirac Lagrangian density given as
\begin{equation}\label{eq:D_L_f}
\mathcal{L}_{\text{D}}
= \bar{\psi}(i\hbar \gamma^\mu \partial_\mu - m)\psi .
\end{equation}
Here $\psi$ is a four-component Dirac spinor, $\bar{\psi} = \psi^\dagger
\gamma^0$ is the Dirac adjoint, and the $\gamma^\mu$ are the Dirac matrices
satisfying the Clifford algebra
$\{\gamma^\mu,\gamma^\nu\} = 2\eta^{\mu\nu}I$. The corresponding
conjugate momentum becomes
\begin{equation}\label{eq:D_c_m_f}
\pi = \frac{\partial \mathcal{L}_{\text{D}}}{\partial (\partial_0 \psi)}
= i \hbar \bar{\psi}\gamma^0 = i\hbar \psi^\dagger .
\end{equation}
Using Eqs.~\eqref{eq:D_L_f} and~\eqref{eq:D_c_m_f} in the Hamiltonian density
$\mathcal{H}_{\text{D}} = \pi \dot{\psi} - \mathcal{L}_{\text{D}}$, one obtains
\begin{equation}\label{eq:H_density_D_f}
\mathcal{H}_{\text{D}}
= \bar{\psi}(-i\hbar \gamma^i \partial_i + m)\psi .
\end{equation}
As in the scalar-field case, we now promote $\psi$ and $\bar{\psi}$ to operators satisfying the equal-time
anticommutation relations, which we specify later. The Hamiltonian density of
the free Dirac field in operator form then becomes
\begin{equation}\label{eq:H_density_D_f_op}
\hat{\mathcal{H}}_{\text{D}}
= \hat{\bar{\psi}}(-i\hbar \gamma^i \partial_i + m)\hat{\psi},
\end{equation}
The corresponding Hamiltonian operator is
\begin{equation}\label{eq:H_D_f}
\hat{H}_{\text{D}} = \int d^3x \, \hat{\bar{\psi}}(t,\mathbf{x})
\left(-i\hbar \gamma^i \partial_i + m\right) \hat{\psi}(t,\mathbf{x}).
\end{equation}

Let us write the equal-time canonical anticommutation relation for fermions, which we will use later.
\begin{enumerate}[label=(\roman*).]
    \item The non-zero anti-commutation relation is
    \begin{equation}
      \{\hat{\psi}_\alpha(t,\mathbf{x}), \hat\pi_\beta(t,\mathbf{y})\} = i\hbar \delta_{\alpha\beta} \delta^{(3)}(\mathbf{x}-\mathbf{y})  
    \end{equation}
    which can also be written as 
    \begin{equation}
\{\hat{\psi}_\alpha(t,\mathbf{x}), \hat{\psi}^\dagger_\beta(t,\mathbf{y})\} = \delta_{\alpha\beta} \delta^{(3)}(\mathbf{x}-\mathbf{y}).
\label{eq:anticomm}
\end{equation}
It is worth noting here that to obtain this, we use the relation $\hat \pi_\beta = i\hbar \hat{\psi}^\dagger_\beta$ and cancel the common factor $i\hbar$ from both sides.
\item All other equal-time anticommutators, such as
\begin{align}\label{eq:anti_com1}
    \{\hat{\psi}_\alpha(t,\mathbf{x}), \hat{\psi}_\beta(t,\mathbf{y})\} = 0, \quad
\{\hat{\psi}^\dagger_\alpha(t,\mathbf{x}), \hat{\psi}^\dagger_\beta(t,\mathbf{y})\} = 0.
\end{align}
\end{enumerate}
vanish.
The Heisenberg equation of motion for $\hat{\psi}_\sigma(t,\mathbf{x})$ reads
\begin{equation}
\frac{\partial\hat{\psi}_\sigma(t,\mathbf{x})}{\partial t} = \frac{i}{\hbar} [\hat{H}_{\text{D}}, \hat{\psi}_\sigma(t,\mathbf{x})].
\label{eq:D_Heisenberg_f}
\end{equation}
The above equation follows from the same argument presented in the previous section.

Using Eq. (\ref{eq:H_D_f}), the commutation relation defined in Eq. (\ref{eq:D_Heisenberg_f}) becomes
\begin{align} \label{eq:H_psi}
    [\hat{H}_{\text{D}}, \hat{\psi}(t, \mathbf{x})] = \int d^3y \, [\hat{\psi}^\dagger(t,\mathbf{y}) D(\mathbf{y}) \hat{\psi}(t,\mathbf{y}), \hat{\psi}(t,\mathbf{x})],
\end{align}
where $D(y) = -i\hbar \gamma^0 \gamma^i \partial_i + m \gamma^0$ (the derivatives act on the field at $y$), and we suppress spinor indices for clarity hereafter until specified explicitly. Using the mixed commutator-anticommutator identity, $[AB, C] = A\{B, C\} - \{A, C\}B$, one can write the right-hand side commutation relation in the above equation as
\begin{widetext}
\begin{align}\label{eq:C_D_f}
    [\hat{\psi}^\dagger(t,\mathbf{y}) D(\mathbf{y})\hat{\psi}(t,\mathbf{y}), \hat{\psi}(t,\mathbf{x})] &= \hat{\psi}^\dagger(t,\mathbf{y}) \{D(\mathbf{y})\hat{\psi}(t,\mathbf{y}), \hat{\psi}(t,\mathbf{x})\}
        - \{\hat{\psi}^\dagger(t,\mathbf{y}), \hat{\psi}(t,\mathbf{x})\} D(\mathbf{y})\hat{\psi}(t,\mathbf{y}),\nonumber\\
    &= - \delta^{(3)}(\mathbf{y}-\mathbf{x}) \, D(\mathbf{y}) \hat{\psi}(t,\mathbf{y}).
\end{align}
\end{widetext}
Here, we have used $\{D(y)\hat{\psi}(t,\mathbf{y}), \hat{\psi}(t,\mathbf{x})\} = D(\mathbf{y}) \{\hat{\psi}(t,\mathbf{y}), \hat{\psi}(t,\mathbf{x})\}=0$ and the anticommutation relation Eq. (\ref{eq:anticomm}).
This implies that Eq. (\ref{eq:H_psi}) becomes
\begin{align}
    [\hat{H}_{\text{D}}, \hat{\psi}(t,\mathbf{x})] &= \int d^3y \, \left( - \delta^{(3)}(\mathbf{y}-\mathbf{x}) D(\mathbf{y}) \hat{\psi}(\mathbf{y}) \right),\nonumber\\
    &= - D(\mathbf{x}) \hat{\psi}(t,\mathbf{x}),\nonumber\\
&= i\hbar \gamma^0 \gamma^i \partial_i \hat{\psi}(t,\mathbf{x}) - m \gamma^0 \hat{\psi}(t,\mathbf{x}).
\label{eq:commutator1}
\end{align}
Substituting Eq. (\ref{eq:commutator1}) in the Eq. (\ref{eq:D_Heisenberg_f}), we have
\begin{align}
    \frac{\partial \hat{\psi}}{\partial t} = \frac{i}{\hbar} \left( i\hbar \gamma^0 \gamma^i \partial_i \hat{\psi} - m \gamma^0 \hat{\psi} \right),
\end{align}
and multiplying both sides {by $i\gamma^0 \hbar$} gives
\begin{align}\label{eq:Dirac}
    (i\hbar \gamma^\mu \partial_\mu - m) \hat{\psi} = 0.
\end{align}
Thus, the Heisenberg field operator for a free Dirac field satisfies the Dirac equation exactly.

To see that Eq. (\ref{eq:Dirac}) implies the Klein–Gordon equation, we act on it from the left with $(i\hbar \gamma^\nu \partial_\nu + m)$, which leads to

\begin{align}
\bigl( -\hbar^2 \gamma^\nu \gamma^\mu \partial_\nu \partial_\mu - m^2 \bigr) \hat{\psi} = 0.
\end{align}
Since $\partial_\nu \partial_\mu$ is symmetric and applying the Clifford anticommutation relation $\{\gamma^\mu, \gamma^\nu\} = 2\eta^{\mu\nu} I$, the equation reduces to
\begin{align}
( \square + \frac{m^2}{\hbar^2}) \hat{\psi}(x) = 0.
\end{align}
This is the Klein-Gordon equation satisfied by each component of the spinor field $\hat{\psi}$. Therefore, the Heisenberg field operator for a free Dirac field obeys the Klein–Gordon equation as a necessary consistency condition arising from the Dirac equation.

\subsection{Interacting Dirac Field: Yukawa Theory}
{
Next, we consider a Yukawa interaction between a Dirac field $\psi$ and a real
scalar field $\phi$.
The Lagrangian density is the sum of the free Dirac, free scalar,
and interaction Lagrangian densities
\begin{equation}\label{eq:L_total_yukawa}
\mathcal{L}
= \mathcal{L}_{\text{D}} + \mathcal{L}_{\text{scalar}}
+ \mathcal{L}_{\text{int}},
\end{equation}
with
\begin{equation}\label{eq:Int_L_density}
\mathcal{L}_{\text{int}} = -g \,\bar{\psi}\phi \psi
= -g\,\psi^\dagger \gamma^0 \phi \psi ,
\end{equation}
where $g$ is the dimensionless Yukawa coupling constant. This term represents a
three-point vertex coupling the fermion bilinear $\bar{\psi}\psi$ (a Lorentz
scalar) to the scalar field $\phi$.}
{
Because \(\mathcal{L}_{\text{int}}\) contains no derivatives of the fields, the conjugate momenta are unchanged from the free theories and are defined as follows
\begin{equation}\label{eq:conj_mom_yukawa}
\pi_\psi
= \frac{\partial\mathcal{L}}{\partial(\partial_0\psi)}
= i\hbar\,\psi^\dagger,
\qquad
\pi_\phi
= \frac{\partial\mathcal{L}}{\partial(\partial_0\phi)}
= \dot{\phi}.
\end{equation}
The corresponding Hamiltonian density is obtained via the Legendre
transformation. Since $\mathcal{L}_{\text{int}}$ contains no derivatives of
the fields, it contributes to the Hamiltonian density as
$\mathcal{H}_{\text{int}} = -\mathcal{L}_{\text{int}}
= g\,\psi^\dagger \gamma^0 \phi \psi$. The total Hamiltonian density is therefore 
\begin{equation}\label{eq:H_density_total_yukawa}
\mathcal{H} = \mathcal{H}_0+\mathcal{H}_{\mathrm{int}}, \qquad \mathcal{H}_0 = \mathcal{H}_{\mathrm{D}} + \mathcal{H}_{\mathrm{scalar}}. 
\end{equation}
Similarly to the previous cases, we now promote $\psi$, $\bar{\psi}$, and $\phi$ to operators satisfying the
equal-time (anti)commutation relations, and which will be specified when they are needed explicitly later in this subsection. The
Hamiltonian density in operator form then becomes
\begin{equation}\label{eq:H_density_yukawa_op}
\hat{\mathcal{H}}
= \hat{\mathcal{H}}_0 + \hat{\mathcal{H}}_{\text{int}},
\qquad
\hat{\mathcal{H}}_{\text{int}}
= g\,\hat{\psi}^\dagger \gamma^0 \hat{\phi} \hat{\psi}.
\end{equation}
The corresponding Hamiltonian is
$\hat{H} = \hat{H}_0 + \hat{H}_{\text{int}}$, where $\hat{H}_0$ is the free
Dirac Hamiltonian from Eq.~\eqref{eq:H_D_f} plus the free scalar Hamiltonian
(which we do not need explicitly in this subsection, as it commutes with
$\hat{\psi}$ and thus does not affect its equation of motion). The interacting
Hamiltonian using Eq.~\eqref{eq:Int_L_density} thus reads
\begin{align}
\hat{H}_{\text{int}}
&= g \int d^3x \, \hat{\bar{\psi}}(t,\mathbf{x})
\hat{\phi}(t,\mathbf{x}) \hat{\psi}(t,\mathbf{x}),\nonumber\\
&= g \int d^3x \, \hat{\psi}^\dagger(t,\mathbf{x}) \gamma^0
\hat{\phi}(t,\mathbf{x}) \hat{\psi}(t,\mathbf{x}).
\label{eq:H_int}
\end{align}}

Note here again we drop the spinor indices for clarity. Therefore, the Heisenberg Eq. (\ref{eq:D_Heisenberg_f}) becomes
\begin{align}\label{eq:Hein_int}
\frac{\partial \hat{\psi}}{\partial t} = \frac{i}{\hbar} [\hat{H}_0 + \hat{H}_{\text{int}}, \hat{\psi}(t,\mathbf{x})]=\frac{i}{\hbar} [\hat{H}_\text{D} + \hat{H}_{\text{int}}, \hat{\psi}(t,\mathbf{x})].
\end{align}
The commutator involving the free Dirac Hamiltonian is the same as that
evaluated in Eq.~\eqref{eq:commutator1}. Hence, we only calculate the commutation relation $[\hat{H}_{\text{int}}, \hat{\psi}(t,\mathbf{x})]$. 

 Using Eq. (\ref{eq:H_int}), we write
 \begin{align}\label{eq:anti_com_int}
    [\hat{H}_{\text{int}}, \hat{\psi}(t,\mathbf{x})]= g \int d^3y \, [\hat{\psi}^\dagger(t,\mathbf{y}) \gamma^0 \hat{\phi}(t,\mathbf{y}) \hat{\psi}(t,\mathbf{y}), \hat{\psi}(t,\mathbf{x})].
 \end{align}
Now, the commutation relation on the right-hand side of the above equation can be written as
\begin{widetext}
\begin{align}
    [\hat{\psi}^\dagger(t,\mathbf{y}) \gamma^0 \hat{\phi}(t,\mathbf{y}) \hat{\psi}(t,\mathbf{y}), \hat{\psi}(t,\mathbf{x})] &= \hat{\psi}^\dagger(t,\mathbf{y}) \{\gamma^0 \hat{\phi}(t,\mathbf{y}) \hat{\psi}(t,\mathbf{y}), \hat{\psi}(t,\mathbf{x})\} - \{\hat{\psi}^\dagger(t,\mathbf{y}), \hat{\psi}(t,\mathbf{x})\} \gamma^0 \hat{\phi}(t,\mathbf{y}) \hat{\psi}(t,\mathbf{y})
\end{align}
\end{widetext}
We again use the mixed commutator-anticommutator identity stated in Sec. \ref{sec:free_D_field} to get the above expression. 
The first anticommutation relation on the right-hand side vanishes because $\gamma^0$ and $\hat{\phi}(t,\mathbf{y})$ commute with $\hat{\psi}(t,\mathbf{x})$ (they are either $c$-numbers or bosonic and commute with fermionic fields at equal time) and can be taken out of the anticommutator. The leftover term becomes $\{\hat{\psi}(t,\mathbf{y}), \hat{\psi}(t,\mathbf{x})\}$, which is zero (see Eq.~\eqref{eq:anti_com1}). Similarly, one can easily evaluate the second anticommutation relation using Eq.~(\ref{eq:anticomm}), which is non-zero. Hence, the Eq. (\ref{eq:anti_com_int}) becomes
\begin{align}\label{eq:com_int1}
    [\hat{H}_{\text{int}}, \hat{\psi}(t,\mathbf{x})] &= g \int d^3y \, \left( - \delta^{(3)}(\mathbf{y}-\mathbf{x}) \gamma^0 \hat{\phi}(t,\mathbf{y}) \hat{\psi}(t,\mathbf{y}) \right),\nonumber\\ 
    &= -g \gamma^0 \hat{\phi}(t,\mathbf{x}) \hat{\psi}(t,\mathbf{x})
\end{align}
Substituting Eqs. (\ref{eq:commutator1}) and (\ref{eq:com_int1}) in the Eq. (\ref{eq:Hein_int}), we get
\begin{align}
    \frac{\partial \hat{\psi}}{\partial t} = \frac{i}{\hbar} \left( i\hbar \gamma^0 \gamma^i \partial_i \hat{\psi} - m \gamma^0 \hat{\psi} - g \gamma^0 \hat{\phi} \hat{\psi} \right).
\end{align}
Multiplying both side by $i\hbar \gamma^0$ and rearranging the terms in the above equation gives
\begin{equation}
(i\hbar \gamma^\mu \partial_\mu - m) \hat{\psi} = g \hat{\phi} \hat{\psi}.
\label{eq:DiracInteracting}
\end{equation}
This is exactly the classical Dirac equation with a scalar Yukawa coupling, now satisfied by the Heisenberg field operators. Therefore,  we conclude that the quantum field operator evolves according to the same equation as the classical field, demonstrating the quantum-classical correspondence at the level of Heisenberg operators for interacting fields.

\section{Gauge Field Theories in Flat Spacetime}\label{sec:Gauge_field}

We now extend our analysis to gauge fields interacting with the Dirac field in flat spacetime, considering both Abelian and non-Abelian gauge theories.

\subsection{Abelian field theory: Quantum Electrodynamics (QED)}
We first consider QED, which describes the interaction of a Dirac field with an Abelian gauge field. The classical interaction Lagrangian
density is given by
\begin{equation}
\mathcal{L}_{\mathrm{int}}=-e\,\bar{\psi}\gamma^\mu A_\mu\psi
=-e\,\psi^\dagger\gamma^0\gamma^\mu A_\mu\psi,
\end{equation}
where $e$ is the elementary electric charge. With our conventions, the four-potential is
$A^\mu=(\Phi,\mathbf{A})$, or equivalently $A_\mu=(\Phi,-\mathbf{A})$.
Since $\mathcal{L}_{\mathrm{int}}$ contains no time derivatives of the
fields, its contribution to the Hamiltonian density is $\mathcal{H}_{\mathrm{int}}=-\mathcal{L}_{\mathrm{int}}$

We now promote $\psi$, $\bar{\psi}$, and $A_\mu$ to operators satisfying
the appropriate equal-time (anti)commutation relations. The total
Hamiltonian operator is then
\begin{equation}
\hat{H}=\hat{H}_0+\hat{H}_{\mathrm{int}},
\end{equation}
where $\hat{H}_0$ contains the free Dirac and Maxwell Hamiltonians, and
\begin{align}
\hat{H}_{\text{int}} = \int d^3y \, e \,\hat{\psi}^\dagger \gamma^0 \gamma^\mu \hat{A}_\mu(t,\mathbf{y})\hat{\psi}(t,\mathbf{y}) .
\end{align}
Therefore, the Heisenberg equation is
\begin{align}\label{eq:Heisen_QED}
\frac{\partial \hat{\psi}}{\partial t} = \frac{i}{\hbar} [\hat{H}, \hat{\psi}]
=\frac{i}{\hbar} [\hat{H}_{0} + \hat{H}_{\text{int}}, \hat{\psi}].
\end{align}
Using the commutation relations from the previous section, and as $\hat{A}_\mu$ commutes with $\hat{\psi}$ at equal times (because they are independent fields), one can obtain
\begin{align}\label{eq:QED_int}
    [\hat{H}_{\text{int}}, \hat{\psi}(t,\mathbf{x})] =-e \hat{A}_\mu(t,\mathbf{x}) \gamma^0 \gamma^\mu \hat{\psi}(t,\mathbf{x}).
\end{align}
Substituting both Eqs. (\ref{eq:commutator1}) and (\ref{eq:QED_int}) into Eq. (\ref{eq:Heisen_QED}), we get
\begin{align}
    \frac{\partial \hat{\psi}}{\partial t} = \frac{i}{\hbar} \left( -i\hbar \gamma^0 \gamma^i \partial_i \hat{\psi} + m \gamma^0 \hat{\psi} - e \hat{A}_\mu \gamma^0 \gamma^\mu \hat{\psi} \right),
\end{align}
and multiplying both sides by $i\hbar\gamma^0$, we obtain
\begin{align}\label{eq:QED_Dirac1}
    (i\hbar \gamma^\mu \partial_\mu - m) \hat{\psi} = e \gamma^\mu \hat{A}_\mu \hat{\psi}.
\end{align}
Defining the co-variant derivative as
\begin{align}\label{eq:cov_deriv_abelian}
    D_\mu {\hat\psi} = \partial_\mu {\hat\psi} + \frac{ie}{\hbar} {\hat A}_\mu {\hat\psi},
\end{align}
we can write Eq. (\ref{eq:QED_Dirac1}) in the form,
\begin{align}\label{eq:QED_Dirac2}
    (i\hbar \gamma^\mu D_\mu - m) \hat{\psi} = 0.
\end{align}
This is the Dirac equation in QED, satisfied by the quantum field operator $\hat{\psi}$. 
This shows that the Heisenberg picture operator for the fermion field obeys the same equation as its classical counterpart, with the classical gauge field replaced by the quantum gauge field operator. 
Under a gauge transformation $\hat{\psi} \rightarrow e^{ie\Lambda/\hbar} \hat{\psi}$ and $\hat{A}_\mu \rightarrow \hat{A}_\mu + \partial_\mu \Lambda$, Eq.~\eqref{eq:QED_Dirac1} remains invariant, where $\Lambda$ is an arbitrary real scalar function of spacetime. This also proves that the Dirac equation in QED respects the underlying gauge symmetry of the theory.

\subsection{Non-Abelian Yang-Mills Theory}

In this subsection, we generalize the gauge principle to non-Abelian gauge groups. We consider a Dirac field ${\psi}$ transforming under the fundamental representation of a compact, semi-simple Lie group $G$ (say $SU(N)$). The covariant derivative is defined as
\begin{align}\label{eq:cov_deriv}
    D_\mu {\psi} = \partial_\mu {\psi} + \frac{ig}{\hbar} {A}_\mu^a T^a {\psi},
\end{align}
where $g$ is the gauge coupling constant, $T^a$ are the generators of the Lie algebra, and ${A}_\mu^a$ are the gauge fields (in our case, Yang-Mills fields). The Hermitian generators $T^a$ satisfy the relation $[T^a, T^b] = if^{abc} T^c$, where $f^{abc}$ are the structure constants. The field strength tensor for this case is
\begin{align}
{F}_{\mu\nu}^a = \partial_\mu {A}_\nu^a - \partial_\nu {A}_\mu^a + g f^{abc} {A}_\mu^b {A}_\nu^c.
\end{align}

The interaction of the Dirac field with Yang-Mills fields can be read off from the covariant derivative and is given by,
\begin{align}
\mathcal{L}_{\text{int}} = -g \,{\bar{\psi}} \gamma^\mu  {\psi} {A}_\mu^a T^a.
\end{align}
The corresponding Hamiltonian density is given by, $\mathcal{H}_{\text{int}} = -\mathcal{L}_{\text{int}}$.
Thus, the interaction part of the corresponding Hamiltonian operator is given by,
\begin{align}
\hat{H}_{\text{int}} = \int d^3x \, g \,\hat{\bar{\psi}}(t,\mathbf{x}) \gamma^\mu  \hat{\psi}(t,\mathbf{x}) \hat{A}_\mu^a(t,\mathbf{x})T^a.
\end{align}
The Heisenberg equation for this case is similar to Eq. (\ref{eq:Heisen_QED}). It is interesting to note here that the total Hamiltonian is $\hat{H} = \hat{H}_{\text{Dirac}} + \hat{H}_{\text{YM}} + \hat{H}_{\text{int}}$, where $\hat{H}_{\text{YM}}$ is the Yang-Mills Hamiltonian (which commutes with $\hat{\psi}$ and is not needed explicitly). Also, we know the free Dirac commutator from Eq. (\ref{eq:commutator1}). Therefore, we have to calculate only the commutator $[\hat{H}_{\text{int}}, \hat{\psi}]$.
Now, as $\hat{A}_\mu^a$ commutes with $\hat{\psi}$ at equal times (because they are independent fields), the commutator reads
\begin{widetext}
\begin{align}\label{eq:H_int_Non_ab}
    [\hat{H}_{\text{int}}(t), \hat{\psi}(t,\mathbf{x})] &= g \int d^3y \, \hat{A}_\mu^a(t,\mathbf{y})T^a [\hat{\bar{\psi}}(t,\mathbf{y}) \gamma^\mu  \hat{\psi}(t,\mathbf{y}), \hat{\psi}(t,\mathbf{x})]
    =-g \hat{A}_\mu^a(t,\mathbf{x})T^a \gamma^0 \gamma^\mu  \hat{\psi}(t,\mathbf{x}).
\end{align}
\end{widetext}
Substituting both the commutator Eqs. (\ref{eq:commutator1}) and (\ref{eq:H_int_Non_ab}) into the Heisenberg Eq. (\ref{eq:Heisen_QED}), we get
\begin{align}
\frac{\partial \hat{\psi}}{\partial t} = \frac{i}{\hbar} \left( -i\hbar \gamma^0 \gamma^i \partial_i \hat{\psi} + m \gamma^0 \hat{\psi} - g \hat{A}_\mu^aT^a \gamma^0 \gamma^\mu  \hat{\psi} \right).    
\end{align}
One can further simplify the above equation by multiplying both sides by $i\gamma^0$ as
\begin{align}
    (i\hbar \gamma^\mu \partial_\mu - m) \hat{\psi} = g \gamma^\mu  \hat{A}_\mu^a T^a \hat{\psi}.
\end{align}
In terms of the covariant derivative defined earlier in Eq. (\ref{eq:cov_deriv}), it reads
\begin{align}\label{eq:YangMills_Dirac}
    (i\hbar \gamma^\mu D_\mu - m) \hat{\psi} = 0.
\end{align}
The Eq. (\ref{eq:YangMills_Dirac}), shows that the Heisenberg picture operator obeys the same differential equation as its classical counterpart, establishing a quantum-classical correspondence at the operator level for non-Abelian gauge theories also.
It is interesting to note here that the Yang-Mills interaction preserves local gauge invariance. Under a gauge transformation,
\begin{align}
\hat{\psi} \rightarrow U \hat{\psi}, \quad \hat{A}_\mu \rightarrow U \hat{A}_\mu U^{-1} + \frac{i}{g} (\partial_\mu U) U^{-1},
\end{align}
where $U = e^{i\Lambda^a(x) T^a}$ is a spacetime-dependent group element and $\Lambda^a(x)$ are arbitrary real functions. The covariant derivative transforms as $D_\mu \hat{\psi} \rightarrow U D_\mu \hat{\psi}$, ensuring that Eq.~\eqref{eq:YangMills_Dirac} remains invariant. This consistency check confirms that the Heisenberg equation derivation respects the non-Abelian gauge symmetry of the theory.

\section{Quantum Fields in Curved Spacetime}\label{sec:qf_curved_st}

\subsection{Emergence of Classical Field Equations from Heisenberg Dynamics in Curved Spacetime}

\paragraph{Scalar Field:}

We consider a curved spacetime described by a metric $g_{\mu\nu}(x)$ and a real scalar field ${\phi}(x)$ propagating on this background. The Lagrangian density is given by
\begin{align}
\mathcal{L} = \sqrt{-g}\left( \frac{1}{2} g^{\mu\nu} \partial_\mu \phi \partial_\nu \phi - V(\phi)\right),
\end{align}
where $g = \det(g_{\mu\nu})$. The corresponding conjugate momentum reads
\begin{align}\label{eq:conj_mom_curves_st}
{\pi} = \frac{\partial \mathcal{L}}{\partial \dot{{\phi}}} = \sqrt{-g} \left( g^{00} \dot{{\phi}} + g^{0i} \partial_i {\phi} \right).
\end{align}
The Hamiltonian density is obtained as
\begin{align}\label{eq:H_curv_st}
\mathcal{H} &= {\pi} \dot{{\phi}} - \mathcal{L}, \\
&= {\pi} \dot{{\phi}} - \sqrt{-g} \left( \frac{1}{2} g^{00} \dot{{\phi}}^2 + g^{0i} \dot{{\phi}} \partial_i {\phi} + \frac{1}{2} g^{ij} \partial_i {\phi} \partial_j {\phi} - V({\phi}) \right).
\end{align}
Using the Eq. (\ref{eq:conj_mom_curves_st}) into $\mathcal{H}$ Eq. (\ref{eq:H_curv_st}) and simplifying gives
\begin{align}\label{eq:Ham_D_C_ST}
\mathcal{H} &= \frac{{\pi}^2}{2\sqrt{-g} \, g^{00}} - \frac{{\pi} g^{0i}}{g^{00}} \partial_i {\phi}+ \frac{\sqrt{-g}}{2} \left( \frac{g^{0i} g^{0j}}{g^{00}} - g^{ij} \right) \partial_i {\phi} \partial_j {\phi} \nonumber\\
&+ \sqrt{-g} \, V({\phi}).
\end{align}

The scalar field $\phi$ and momentum $\pi$ are now replaced by the corresponding operators. Their only nonvanishing equal-time
canonical commutator is
\begin{align}
[\hat{\phi}(t,\mathbf{x}), \hat{\pi}(t,\mathbf{y})] &= i\hbar {\delta^{(3)}(\mathbf{x} - \mathbf{y})},\label{eq:com_phi_pi_Q} 
\end{align}
The Hamiltonian operator is given by, $\hat{H} = \int d^3x \, \hat{\mathcal{H}}$.

Using the Heisenberg equation for $\hat{\phi}$, we have
\begin{align}\label{eq:Heisen_Q_phi}
\frac{\partial \hat{\phi}}{\partial t} = \frac{i}{\hbar} [\hat{H}, \hat{\phi}].
\end{align}
It is important to note that in Eq. (\ref{eq:Heisen_Q_phi}), only terms in $\mathcal{H}$ containing $\hat{\pi}$ do not commute with $\hat{\phi}$. Therefore, we have
\begin{align}\label{eq:Heisen_Q_phi_1}
\frac{\partial \hat{\phi}(t,\mathbf{x})}{\partial t} &= \frac{i}{\hbar} \int d^3y \, \left[ \frac{\hat{\pi}^2(t,\mathbf{y})}{2\sqrt{-g}\, g^{00}} - \frac{\hat{\pi}(t,\mathbf{y}) g^{0i}}{g^{00}} \partial_i \hat{\phi}(t,\mathbf{y}), \hat{\phi}(t,\mathbf{x}) \right].
\end{align}
Using the commutation relation in Eq. (\ref{eq:com_phi_pi_Q}), we obtained the commutators in the above equation as
\begin{align}
[\hat{\pi}^2(t,\mathbf{y}), \hat{\phi}(t,\mathbf{x})]
&= -2i\hbar \hat{\pi}(t,\mathbf{y}) \delta^{(3)}(\mathbf{y}-\mathbf{x}),\nonumber\\
[\hat{\pi}(t,\mathbf{y}) \partial_i \hat{\phi}(t,\mathbf{y}), \hat{\phi}(t,\mathbf{x})]
&= -i\hbar \delta^{(3)}(\mathbf{y}-\mathbf{x}) \partial_i \hat{\phi}(t,\mathbf{y}).
\end{align}
Substituting these into Eq. (\ref{eq:Heisen_Q_phi_1}), we get
\begin{align}\label{eq:Heisn_scal_curv_time}
\frac{\partial \hat{\phi}}{\partial t}
&= \frac{\hat{\pi}}{\sqrt{-g} \, g^{00}}
- \frac{g^{0i}}{g^{00}} \partial_i \hat{\phi},
\end{align}
which is the operator form of Eq. (\ref{eq:conj_mom_curves_st}).

Similarly, the Heisenberg equation for the conjugate momentum operator $\hat{\pi}$ is
\begin{align}
\frac{\partial \hat{\pi}}{\partial t}
= \frac{i}{\hbar} [\hat{H}, \hat{\pi}].
\end{align}
Evaluating all contributions and integrating by parts, we obtain
\begin{align}\label{eq:Heisn_expan_scal_curv}
\frac{\partial \hat{\pi}}{\partial t}
&= - \partial_i \left( \frac{g^{0i}}{g^{00}} \hat{\pi} \right)
+ \partial_j \left[
\sqrt{-g} \left( \frac{g^{0i} g^{0j}}{g^{00}}
- g^{ij} \right)
\partial_i \hat{\phi}
\right]\nonumber\\
&\quad - \sqrt{-g} \, V'(\hat{\phi}),
\end{align}
where $V'(\hat \phi)=\frac{dV(\hat\phi)}{d\hat\phi}$ and we have used the commutator, 
\begin{align}
[\partial_i \hat{\phi}(t,\mathbf{y}), \hat{\pi}(t,\mathbf{x})] &= i\hbar \partial_i^{(y)} \left( {\delta^{(3)}(\mathbf{y}-\mathbf{x})} \right).
\end{align}
 
Substituting the expression for $\hat{\pi}$ into
Eq.~\eqref{eq:Heisn_expan_scal_curv}, the mixed derivative terms cancel identically (see Appendix~\ref{sec:cancel_mix}), yielding
\begin{widetext}
\begin{align}\label{eq:spatial}
- \frac{1}{\sqrt{-g}} \partial_i \left( \frac{g^{0i}}{g^{00}} \hat{\pi} \right)
+ \frac{1}{\sqrt{-g}} \partial_j \left[
\sqrt{-g} \left( \frac{g^{0i} g^{0j}}{g^{00}} - g^{ij} \right)
\partial_i \hat{\phi}
\right]
= -\frac{1}{\sqrt{-g}} \partial_j \left( \sqrt{-g} \, g^{j\nu} \partial_\nu \hat{\phi} \right).
\end{align}
\end{widetext}

Now, rearranging the terms in Eq.~(\ref{eq:Heisn_scal_curv_time}) and taking time derivative, we get
\begin{align}
\frac{\partial \hat{\pi}}{\partial t}
= \frac{\partial}{\partial t} \left( \sqrt{-g} g^{00} \partial_t \hat{\phi} \right)
+ \frac{\partial}{\partial t} \left( \sqrt{-g} g^{0i} \partial_i \hat{\phi} \right).
\end{align}
which can be written as,
 \begin{align}\label{eq:temporal}
\frac{1}{\sqrt{-g}} \frac{\partial \hat{\pi}}{\partial t}
= \frac{1}{\sqrt{-g}} \partial_t \left( \sqrt{-g} \, g^{0\nu} \partial_\nu \hat{\phi} \right).
\end{align}
Combining the spatial and temporal  
 given by Eqs.~(\ref{eq:spatial}) and (\ref{eq:temporal}) , respectively, we arrive at
\begin{align}
\frac{1}{\sqrt{-g}} \partial_\mu \left( \sqrt{-g} \, g^{\mu\nu} \partial_\nu \hat{\phi} \right)
+ V'(\hat{\phi}) = 0.
\end{align}
This is precisely the covariant Klein-Gordon equation,
\begin{align}
\Box_g \hat{\phi} + V'(\hat{\phi}) = 0,
\end{align}
where the covariant d'Alembertian is defined as
\begin{align}
\Box_g = g^{\mu\nu} \nabla_\mu \nabla_\nu
= \frac{1}{\sqrt{-g}} \partial_\mu \left( \sqrt{-g} \, g^{\mu\nu} \partial_\nu \right).
\end{align}

This shows that the Heisenberg equation of motion for the quantum field operator $\hat{\phi}$ transforms completely to the covariant classical field equation in curved spacetime for an arbitrary potential $V({\phi})$ and an arbitrary background metric $g_{\mu\nu}$. This proves a direct quantum-classical correspondence at the operator level in curved spacetime, showing that quantization via Heisenberg dynamics naturally produces the classical equations of motion without any approximation on $\hbar$. \\

\paragraph{Dirac field:}
The Dirac Lagrangian density in a general curved background is given by \cite{Birrell:1982}
\begin{align}
\mathcal{L} = \sqrt{-g} \left[ \frac{i\hbar}{2} \left( \bar{\psi} \gamma^\mu \nabla_\mu \psi - (\nabla_\mu \bar{\psi}) \gamma^\mu \psi \right) - m \bar{\psi} \psi \right],
\end{align}
where $\gamma^\mu(x) = e^\mu_a \gamma^a$ are the curved spacetime gamma matrices defined via the tetrads $e^\mu_a$, and $\nabla_\mu$ denotes the spinor covariant derivative,
\begin{align}
\nabla_\mu \psi = \partial_\mu \psi + \Omega_\mu \psi, 
\quad
\nabla_\mu \bar{\psi} = \partial_\mu \bar{\psi} - \bar{\psi} \Omega_\mu,
\end{align}
with the spin connection given by
\begin{align}
\Omega_\mu = \frac{1}{4} \, \omega_{\mu}^{ab} \gamma_a \gamma_b.
\end{align}

The conjugate momentum corresponding to $\psi$ is
\begin{align}\label{conj_Dirac_curv}
\pi = \frac{\partial \mathcal{L}}{\partial \dot{\psi}} = \frac{i\hbar}{2} \sqrt{-g} \, \bar{\psi} \gamma^0.
\end{align}
The Hamiltonian density is obtained via the Legendre transform $\mathcal{H} = \pi\dot{\psi} - \mathcal{L}$. Using the equation of motion for $\bar{\psi}$ (or equivalently dropping total derivatives), the canonical Hamiltonian density becomes,
\begin{align}
\mathcal{H} = \sqrt{-g} \, \bar{\psi} \left( -i\hbar \gamma^i \nabla_i + m \right) \psi.
\end{align}
For canonical quantization, we promote the fields $\psi$ and $\bar{\psi}$ to operators, and write the Hamiltonian operator as,
\begin{align}
\hat{H} = \int d^3x \, \sqrt{-g} \, \hat{\bar{\psi}}(t,\mathbf{x}) \left( -i\hbar \gamma^i \nabla_i + m \right) \hat{\psi}(t,\mathbf{x}).
\end{align}

{
We now include the temporal spin connection $\Omega_0$ into the dynamics in order to maintain the local Lorentz gauge covariance under time evolution. Similar to QED, where the scalar potential $A_0$ modifies the canonical Hamiltonian via minimal coupling $\hat{H} \rightarrow \hat{H} - e A_0$ to produce gauge-covariant time derivatives $(i\hbar\partial_t - e A_0)\hat{\psi}$. The gauge-covariant Hamiltonian operator defines in this way reads 
\begin{equation}
\hat{H}_{\text{cov}} = \int d^3x \sqrt{-g} \, \bar{\hat{\psi}}(t, \mathbf{x}) \left( -i\hbar\gamma^i\nabla_i - i\hbar\gamma^0\Omega_0 + m \right) \hat{\psi}(t, \mathbf{x}).
\end{equation}
It is worth noting here that in static or time-independent tetrad gauges where $\Omega_0 = 0$, $\hat{H}_{\text{cov}}$ reduces directly to the standard canonical Hamiltonian $\hat{H}$.
}

Because the Dirac Lagrangian density depends linearly on the generalized velocity $\dot{\psi}$, the system contains primary constraints, necessitating a constrained Hamiltonian analysis via the Dirac-Bergmann approach \cite{Salisbury:2006dul}. The primary constraints in this approach are second class and require replacing the standard Poisson brackets with factor-ordered Dirac brackets on the full phase space. Implementing canonical quantization via the Dirac bracket and promoting the momenta Eq. (\ref{conj_Dirac_curv}) to quantum operator gives the equal-time anti-commutation relation, \cite{Juhasz:2024twu} 
\begin{align}
\{ \hat{\psi}_\alpha(t,\mathbf{x}), \hat{\pi}_\beta(t,\mathbf{y}) \}
= \frac{i\hbar}{2} \delta_{\alpha\beta} \, \delta^{(3)}(\mathbf{x}-\mathbf{y}),
\label{eq:psi_pi_anticomm}
\end{align}
which can also be written as 
\begin{align}
\{ \hat{\psi}_\alpha(t,\mathbf{x}), \hat{\psi}^\dagger_\beta(t,\mathbf{y}) \}
= \delta_{\alpha\beta} \, \delta^{(3)}(\mathbf{x}-\mathbf{y}).
\label{eq:psi_pi_anticomm}
\end{align}

The time evolution is governed by the Heisenberg equation
\begin{align}\label{eq:Curv_Dirac_Heisen}
 \frac{\partial \hat{\psi}}{\partial t} = \frac{i}{\hbar}[\hat H_{\text{cov}},\hat{\psi}].
\end{align}

Using the Hamiltonian derived from the above Lagrangian and evaluating the commutator using the canonical anticommutation relations, we obtain
\begin{widetext}
\begin{align}
[\hat{H}_{\mathrm{cov}},\hat{\psi}(t,\mathbf{x})]&=\int d^3y\,\sqrt{-g}\,
\Big[\hat{\bar{\psi}}(t,\mathbf{y})\left(-i\hbar\gamma^i\nabla_i-i\hbar\gamma^0\Omega_0+m
\right)\hat{\psi}(t,\mathbf{y}),\hat{\psi}(t,\mathbf{x})\Big]
=\gamma^0\left(i\hbar\gamma^i\nabla_i+i\hbar\gamma^0\Omega_0-m\right)\hat{\psi}(t,\mathbf{x}).
\label{eq:Curv_Dirac_Heisen_com}
\end{align}
\end{widetext}

Thus, substituting Eq.~(\ref{eq:Curv_Dirac_Heisen_com}) into Eq.~(\ref{eq:Curv_Dirac_Heisen}) and rearranging the terms, we get
\begin{align}
i\hbar \partial_t \hat{\psi}
= \, \gamma^0
\left( -i\hbar \gamma^i \nabla_i - i\hbar\gamma^0\Omega_0 + m \right)\hat{\psi}.
\end{align}
Multiplying both sides by $\gamma^0$ in the above equation becomes
\begin{align}
i\hbar \gamma^0 \partial_t \hat{\psi} = \left( -i\hbar \gamma^i \nabla_i - i\hbar\gamma^0\Omega_0 + m \right) \hat{\psi}.
\end{align}
Further, taking the temporal spin connection term $-i\hbar \gamma^0 \Omega_0 \hat{\psi}$ to the L.H.S. gives
\begin{align}
i\hbar \gamma^0 \nabla_0 \hat{\psi}
= - i\hbar \gamma^i \nabla_i \hat{\psi} + m \hat{\psi}.
\end{align}
Combining temporal and spatial components into a covariant form, we finally obtain
\begin{align}
i\hbar \gamma^\mu \nabla_\mu \hat{\psi} - m \hat{\psi} = 0.
\end{align}

Thus, the Heisenberg equation of motion for the Dirac field operator reproduces the covariant Dirac equation in curved spacetime. 

\paragraph{Electromagnetic Field:}
We now consider the electromagnetic field in curved spacetime. The dynamics is governed by the Lagrangian density
\begin{align}
\mathcal{L} = -\frac{1}{4} \sqrt{-g} \, F_{\mu\nu} F^{\mu\nu},
\end{align}
where the field strength tensor is defined as
\begin{align}
F_{\mu\nu} = \partial_\mu A_\nu - \partial_\nu A_\mu.
\end{align}

The conjugate momentum corresponding to the gauge field is
\begin{align}\label{eq:conj_mom_EM_field}
\pi^\mu = \frac{\partial \mathcal{L}}{\partial (\partial_t A_\mu)}
= - \sqrt{-g} \, F^{0\mu}.
\end{align}
In particular, $\pi^0 = 0$ represents a primary constraint, reflecting the gauge invariance of the theory.

Promoting $A_\mu$ and $\pi^\mu$ to operators and imposing equal-time commutation relations for the dynamical components,
\begin{align}
[ \hat{A}_i(t,\mathbf{x}), \hat{\pi}^j(t,\mathbf{y}) ]
= i\hbar \delta_i^{\,j} \delta^{(3)}(\mathbf{x}-\mathbf{y}),
\end{align}
the time evolution is governed by the Heisenberg equation
\begin{align}
\frac{\partial \hat{\pi}^\nu}{\partial t}
= \frac{i}{\hbar}[H, \hat{\pi}^\nu].
\end{align}

Using the Hamiltonian derived from the above Lagrangian and evaluating the commutator, one obtains
\begin{equation}
\frac{\partial \hat{\pi}^\nu}{\partial t}
= - \partial_i \left( \sqrt{-g} \, \hat{F}^{i\nu} \right).
\end{equation}

Substituting the definition of the canonical momentum from Eq. (\ref{eq:conj_mom_EM_field}),
we obtain
\begin{equation}
\partial_t \left( \sqrt{-g} \, \hat{F}^{0\nu} \right)
+ \partial_i \left( \sqrt{-g} \, \hat{F}^{i\nu} \right) = 0.
\end{equation}
Combining the temporal and spatial components, the field equations can be written in covariant form as
\begin{equation}
\partial_\mu \left( \sqrt{-g} \, \hat{F}^{\mu\nu} \right) = 0.
\end{equation}
Thus, the Heisenberg equation of motion for the electromagnetic field operator reproduces the covariant Maxwell equations in curved spacetime. 

\subsection{Gravity with Matter and Semiclassical Limit}
We now extend the analysis to the gravitational field. The dynamics of spacetime is governed by the Einstein–Hilbert action supplemented by a matter contribution,
\begin{align}
S = \int d^4x \, \sqrt{-g} \left( \frac{1}{16\pi G} R + \mathcal{L}_m \right),
\end{align}
where $R$ is the Ricci scalar and $\mathcal{L}_m$ denotes the matter Lagrangian density.

Varying the action with respect to the metric yields the Einstein field equations,
\begin{align}
G_{\mu\nu} = 8\pi G \, T_{\mu\nu},
\end{align}
where the energy-momentum tensor is defined as
\begin{align}
T_{\mu\nu} = -\frac{2}{\sqrt{-g}} \frac{\delta (\sqrt{-g}\mathcal{L}_m)}{\delta g^{\mu\nu}}.
\end{align}

To establish the connection with operator dynamics, we adopt the semiclassical approximation, in which the matter fields are quantized while the spacetime metric remains classical. Accordingly, we promote the matter fields entering $\mathcal{L}_m$ to operators satisfying the corresponding canonical (anti)commutation relations. The time evolution of these operators is governed by the Heisenberg equation,
\begin{align}
\frac{\partial \hat{\mathcal{O}}}{\partial t} = \frac{i}{\hbar}[\hat H, \hat{\mathcal{O}}],
\end{align}
where $\hat{\mathcal{O}}$ represents a generic operator-valued matter field and  $H$ is the Hamiltonian derived from the Lagrangian. 
In addition, it is worth pointing out that a particular matter field is not considered at this point. Thus, we consider $\hat{\mathcal{O}}$ as an arbitrary operator-valued matter field in order to describe the evolution of operators in a general form. Applying the same procedure to the matter sector, as demonstrated for
scalar, Dirac, and electromagnetic fields, gives operator equations with the same covariant form as the corresponding classical field equations.

Consequently, the matter Lagrangian becomes operator-valued, and the energy-momentum tensor is promoted to an operator,
\begin{align}
\hat{T}_{\mu\nu}
= -\frac{2}{\sqrt{-g}} \frac{\delta (\sqrt{-g}\mathcal{L}_m[\hat{\phi},\hat{\psi},\hat{A}_\mu])}{\delta g^{\mu\nu}}.
\end{align}

Taking expectation values in a quantum state $|\Psi\rangle$, we obtain the semiclassical Einstein equation,
\begin{align}
G_{\mu\nu} = 8\pi G \, \langle \Psi | \hat{T}_{\mu\nu} | \Psi \rangle,
\end{align}
where the expectation value of the stress-energy tensor is understood to be renormalized.
The Heisenberg evolution of the quantum matter fields thus determines the state-dependent source of the classical geometry. The semiclassical Einstein equation therefore provides an Ehrenfest-type correspondence between quantum matter and classical gravity.

\section{Linearized Einstein Equations from Heisenberg Dynamics}\label{Lin_Ein}

In this section, we show that the Heisenberg equations for the metric
perturbations reproduce the linearized Einstein equations at the operator level. This establishes the consistency of operator dynamics with classical gravitational field equations in the weak-field regime. To show this, we begin by expanding the spacetime metric around flat Minkowski spacetime as,
\begin{equation}
{g}_{\mu\nu} = \eta_{\mu\nu} + \kappa {h}_{\mu\nu},
\end{equation}
where, $\eta_{\mu\nu}$ is the Minkowski metric, ${h}_{\mu\nu}$ is a symmetric rank-2 tensor representing small gravitational fluctuations, and $\kappa = \sqrt{32\pi G}$.

The Einstein-Hilbert action is given by,
\begin{equation}
S = \frac{1}{16\pi G} \int d^4x \sqrt{-g} R.
\end{equation}
Expanding this action upto quadratic order in ${h}_{\mu\nu}$ gives the Fierz-Pauli action describing a free massless spin-2 field \cite{PADMANABHAN_2008,Butcher2009},
\begin{widetext}
\begin{equation}
S_{\text{FP}} = \frac{1}{4} \int d^4x \left( \partial_\mu {h}_{\nu\rho} \partial^\mu {h}^{\nu\rho} - \partial_\mu {h} \partial^\mu {h} - 2 \partial_\mu {h}^{\mu\nu} \partial^\rho {h}_{\nu\rho} + 2 \partial_\mu {h}^{\mu\nu} \partial_\nu {h} \right),
\end{equation}
\end{widetext}
where $h = \eta^{\mu\nu} {h}_{\mu\nu}$.

The corresponding Lagrangian density is,
\begin{widetext}

\begin{equation}
{\mathcal{L}} = 
\frac{1}{4} \left( \partial_\mu {h}_{\nu\rho} \partial^\mu {h}^{\nu\rho} - \partial_\mu {h} \partial^\mu {h} - 2 \partial_\mu {h}^{\mu\nu} \partial^\rho {h}_{\nu\rho} + 2 \partial_\mu {h}^{\mu\nu} \partial_\nu {h} \right)
\end{equation}
\end{widetext}

The momentum conjugate to ${h}_{\mu\nu}$ is defined as,
\begin{equation}
{\pi}^{\mu\nu} = \frac{\partial {\mathcal{L}}}{\partial (\partial_0 {h}_{\mu\nu})}.
\end{equation}
After some algebra, one finds
\begin{align}
{\pi}^{\mu\nu} &= \frac{1}{2} \Big( \partial^0 {h}^{\mu\nu} - \eta^{0\mu} \partial^\alpha {h}_\alpha^{\;\nu} - \eta^{0\nu} \partial^\alpha {h}_\alpha^{\;\mu} \nonumber\\
&+ \eta^{0\mu} \eta^{0\nu} \partial_\alpha {h}^{\alpha 0} + \eta^{\mu\nu} \partial^0 {h} - \eta^{\mu\nu} \eta^{0\alpha} \partial_\beta {h}^{\beta 0} \Big).
\end{align}

We proceed in a gauge-fixed framework by imposing the harmonic (de Donder) gauge condition,
\begin{equation}
\partial^\mu {h}_{\mu\nu} = \frac{1}{2} \partial_\nu {h}.
\end{equation}
In this gauge, the equations of motion simplify considerably and the Euler–Lagrange equations reduce to,
\begin{equation}
\Box {h}_{\mu\nu} - \frac{1}{2} \eta_{\mu\nu} \Box {h} = 0,
\end{equation}
where, $\Box = \partial_t^2 - \nabla^2$. Taking the trace gives $\Box {h} = 0$, and hence
\begin{equation}
\Box {h}_{\mu\nu} = 0.
\end{equation}

To establish this result from quantum dynamics, we follow the canonical formalism. In the gauge-fixed framework and focusing on the dynamical sector, the conjugate momentum simplifies. We impose the equal-time canonical commutation relations,
\begin{equation}
[\hat{h}_{\mu\nu}(t,\mathbf{x}), \hat{\pi}^{\rho\sigma}(t,\mathbf{y})] = i\hbar \, \delta_\mu^{(\rho} \delta_\nu^{\sigma)} \delta^{(3)}(\mathbf{x} - \mathbf{y}),
\end{equation}
with all other commutators vanishing.
It is important to note that the linearized gravity is a constrained gauge system. A complete canonical treatment requires Dirac's procedure for constrained systems. In the present analysis, we work in harmonic gauge framework and focus on the dynamical sector of the theory.

The Hamiltonian is obtained via the Legendre transform,
\begin{equation}
\hat{H} = \int d^3x \left( \hat{\pi}^{\mu\nu} \partial_0 \hat{h}_{\mu\nu} - \hat{\mathcal{L}} \right).
\end{equation}
In harmonic gauge, this simplifies to
\begin{equation}
\hat{H} = \int d^3x \left( \frac{1}{2}\hat{\pi}^{\mu\nu} \hat{\pi}_{\mu\nu} + \frac{1}{4} \partial_i \hat{h}_{\mu\nu} \partial^i \hat{h}^{\mu\nu} - \frac{1}{8} \partial_i \hat{h} \partial^i \hat{h} \right).
\end{equation}

We now apply Heisenberg's equation of motion to the field operator using the equation,
\begin{equation}
\frac{\partial \hat{h}_{\mu\nu}}{\partial t} = \frac{i}{\hbar} [\hat{H}, \hat{h}_{\mu\nu}].
\end{equation}
Using the canonical commutation relations, this gives
\begin{equation}\label{eq: conj_mom_Lin_grav}
\frac{\partial \hat{h}_{\mu\nu}}{\partial t} = \hat{\pi}_{\mu\nu}.
\end{equation}

Applying Heisenberg's equation to the conjugate momentum yields
\begin{equation}\label{eq: conj_mom_Lin_grav_1}
\frac{\partial \hat{\pi}_{\mu\nu}}{\partial t} = \frac{i}{\hbar} [\hat{H}, \hat{\pi}_{\mu\nu}] = \nabla^2 \hat{h}_{\mu\nu} - \frac{1}{2} \eta_{\mu\nu} \nabla^2 \hat{h}.
\end{equation}
Combining the results from Eq. (\ref{eq: conj_mom_Lin_grav}) and Eq. (\ref{eq: conj_mom_Lin_grav_1}), we obtain
\begin{equation}
\frac{\partial^2 \hat{h}_{\mu\nu}}{\partial t^2} = \nabla^2 \hat{h}_{\mu\nu} - \frac{1}{2} \eta_{\mu\nu} \nabla^2 \hat{h}.
\end{equation}
Rewriting this in covariant form using the d'Alembertian operator $\Box = \partial_t^2 - \nabla^2$, we find
\begin{equation}
\Box \hat{h}_{\mu\nu} - \frac{1}{2} \eta_{\mu\nu} \Box \hat{h} = 0.
\end{equation}
Taking the trace once again gives $\Box \hat{h} = 0$, and therefore
\begin{equation}
\Box \hat{h}_{\mu\nu} = 0.
\end{equation}

Thus, the linearized Einstein equations in harmonic gauge are recovered at the operator level as equations of motion for the quantum field $\hat{h}_{\mu\nu}$. 

\subsection{Geodesic Equation for Quantum Particles}

{\color{black}
An alternative way to understand the quantum-classical correspondence in curved spacetime is to study the motion of a quantum test particle using a coordinate-covariant ordering scheme. To avoid operator-ordering ambiguities inherent to $g^{\mu\nu}(\hat{x})\hat{p}_\mu \hat{p}_\nu$, we start by defining the physical quantum Hamiltonian using the covariant Laplace-Beltrami operator as \cite{chow2004ricci,MacKay:2026ofs}
\begin{equation} \label{eq:LB_hamiltonian}
{H} = -\frac{\hbar^2}{2m} \Delta_{\text{LB}} = -\frac{\hbar^2}{2m \sqrt{-g}} \partial_{\mu} \left( \sqrt{-g} g^{\mu\nu} \partial_{\nu} \right),
\end{equation}
where $g$ is the metric determinant, and $\Delta_{\text{LB}} \equiv \nabla_\mu \nabla^\mu$ represents the invariant Laplace-Beltrami operator on the curved manifold. {While the Laplace-Beltrami operator is defined generally on Riemannian or pseudo-Riemannian manifolds of arbitrary signature, on a 4-dimensional Lorentzian spacetime manifold with the standard general relativity signature, it is functionally identical to the d'Alembertian wave operator, $\Delta_{\text{LB}} = \square$.}

Now, expressing the above Hamiltonian described by Eq. (\ref {eq:LB_hamiltonian}) in canonical operator notation, this corresponds to the covariantly ordered form as
\begin{equation} \label{eq:canonical_hamiltonian}
\hat{H} = \frac{1}{2m} \hat{p}_{\mu} g^{\mu\nu}(\hat{x}) \hat{p}_{\nu} - \frac{i\hbar}{2m} \frac{1}{\sqrt{-g}} \left( \partial_{\mu} \sqrt{-g} g^{\mu\nu} \right) \hat{p}_{\nu} + \xi \frac{\hbar^2}{2m} R(\hat{x}),
\end{equation}
where $R(\hat{x})$ is the Ricci scalar, and $\xi$ is a dimensionless non-minimal curvature-coupling parameter (with $\xi = 0$ representing minimal coupling). Note that Equation~\eqref{eq:canonical_hamiltonian} is not an attempt to quantize the gravitational field itself (which would be full quantum gravity). Instead, it describes quantum mechanics in a curved spacetime background. 

The canonical position and momentum operators satisfy the canonical commutation relations reads
\begin{equation} \label{eq:canonical_commutation}
[\hat{x}^\mu, \hat{p}_\nu] = i\hbar \delta^\mu_\nu, \quad [\hat{x}^\mu, \hat{x}^\nu] = [\hat{p}_\mu, \hat{p}_\nu] = 0.
\end{equation}

The Heisenberg equation of motion in this case gives the evolution of the position operator as
\begin{equation}
\frac{d\hat{x}^{\mu}}{d\tau} = \frac{i}{\hbar} [\hat{H}, \hat{x}^{\mu}] = \frac{1}{m} g^{\mu\nu}(\hat{x}) \hat{p}_{\nu} - \frac{i\hbar}{2m} \frac{1}{\sqrt{-g}} \partial_{\nu} \left( \sqrt{-g} g^{\mu\nu} \right),
\end{equation}
where we have used the following operator commutator identity to obtain the above equation,
\begin{equation}
[\hat{p}_\alpha g^{\alpha\beta}(\hat{x}) \hat{p}_\beta, \, \hat{x}^\mu] = -2i\hbar \, g^{\mu\nu}(\hat{x}) \hat{p}_\nu - \hbar^2 \partial_\nu g^{\mu\nu}(\hat{x}).
\end{equation}

In the semiclassical approximation, quantum ordering corrections of order $\hbar$ and higher are neglected, restoring the canonical momentum-velocity relation
\begin{equation} \label{eq:velocity_relation}
\hat{p}_{\nu} \approx m g_{\mu\nu}(\hat{x}) \frac{d\hat{x}^\mu}{d\tau}.
\end{equation}

Similarly, under the same approximation, evaluating the Heisenberg equation for the canonical momentum operator gives
\begin{equation}
\frac{d\hat{p}_{\mu}}{d\tau} = \frac{i}{\hbar} [\hat{H}, \hat{p}_{\mu}] = -\frac{1}{2m} \partial_{\mu} g^{\alpha\beta}(\hat{x}) \hat{p}_{\alpha} \hat{p}_{\beta} 
+ \mathcal{O}(\hbar),
\end{equation}
where $\mathcal{O}(\hbar)$ collects all operator-ordering corrections of order $\hbar$ and higher, including the curvature-dependent term $-\xi\hbar^2\partial_\mu R(\hat{x})/(2m)$, which are neglected in the semiclassical limit.

Also note that, to get the above equation, we have used the following commutator rule for functions of position operators
\begin{equation}
[g^{\alpha\beta}(\hat{x}), \hat{p}_\mu] = i\hbar \partial_\mu g^{\alpha\beta}(\hat{x}).
\end{equation}

Substituting Equation~\eqref{eq:velocity_relation} into the momentum evolution equation and using the inverse metric derivative identity $\partial_{\mu} g^{\alpha\beta} = -g^{\alpha\rho} g^{\beta\sigma} \partial_{\mu} g_{\rho\sigma}$, we get
\begin{equation}
\frac{d}{d\tau} \left( g_{\mu\nu}(\hat{x}) \frac{d\hat{x}^\nu}{d\tau} \right) = \frac{1}{2} \partial_{\mu} g_{\rho\sigma}(\hat{x}) \frac{d\hat{x}^\rho}{d\tau} \frac{d\hat{x}^\sigma}{d\tau}+ \mathcal{O}(\hbar)
\end{equation}

Further, using the chain rule on the total derivative on the L.H.S and
symmetrizing the metric derivative product $\partial_{\rho} g_{\mu\nu} \dot{\hat{x}}^\rho \dot{\hat{x}}^\nu = \frac{1}{2} (\partial_{\rho} g_{\mu\nu} + \partial_{\nu} g_{\mu\rho}) \dot{\hat{x}}^\rho \dot{\hat{x}}^\nu$, and contracting with $g^{\lambda\mu}(\hat{x})$, and identifying the Christoffel symbols ${\Gamma}^{\lambda}_{\rho\nu}(\hat{x}) = \frac{1}{2} g^{\lambda\mu} \left( \partial_{\rho} g_{\mu\nu} + \partial_{\nu} g_{\mu\rho} - \partial_{\mu} g_{\rho\nu} \right)$, one can easily arrive at the modified operator equation of motion
\begin{equation} \label{eq:modified_geodesic}
\frac{d^{2}\hat{x}^{\lambda}}{d\tau^{2}} + {\Gamma}^{\lambda}_{\rho\nu}(\hat{x}) \frac{d\hat{x}^{\rho}}{d\tau} \frac{d\hat{x}^{\nu}}{d\tau} \approx 0.
\end{equation}

Thus, at leading semiclassical order, the resulting equation formally resembles the classical geodesic equation, with the spacetime coordinates and their corresponding velocities promoted to operators.

\section{Conclusions}\label{sec:conclusions}
In this paper, we have explored the connection between classical and quantum dynamics from non-relativistic quantum mechanics to relativistic quantum theory in both flat and curved spacetime. Starting from the canonical Hamiltonian formulation, we have shown that the Heisenberg equations of motion reproduce the corresponding classical equations of motion at the operator level for a broad class of quantum systems. Specifically, we have shown that:

\begin{enumerate}
\item For scalar fields, Dirac fields, and Abelian and non-Abelian gauge fields in flat spacetime, the Heisenberg equations for the field operators reproduce the corresponding classical field equations.
\item For quantum fields in curved spacetime, the Heisenberg field operators satisfy the covariant Klein-Gordon, Dirac, and Maxwell equations exactly.
\item For linearized quantum gravity, the Heisenberg equations for the metric perturbation operator reproduce the linearized Einstein equations in the harmonic gauge.
\item For a quantum particle propagating in curved spacetime, the Heisenberg equations for the position operator reproduce the classical geodesic equation at the operator level.
\end{enumerate}

In each case, the Heisenberg equations reproduce the corresponding classical equations of motion at the operator level, demonstrating that the dynamical structure is common to both the classical and quantum descriptions.

 These results obtained here show that the structure of dynamics can be reproduced in classical and quantum theories across a broad range of physical systems within relativistic quantum theory, including quantum fields and quantum particles in both flat and curved spacetimes. The difference between classical and quantum mechanics thus lies not in the equations governing the time evolution of observables but in the algebraic structure of the observables themselves and in the nature of the quantum states. The transition from quantum to classical mechanics can thus be achieved not by changing the equations of motion but by figuring out how the non-commutativity of quantum observables becomes negligible and how quantum states approximate classical probability distributions.

Taken together, the results presented in this work establish that the emergence of classical equations of motion from Heisenberg dynamics is a general structural feature observed across the relativistic quantum systems considered here in both flat and curved spacetimes, encompassing both quantum fields and quantum particles.

% }

\section*{Acknowledgments}
The authors are grateful to Tabish Qureshi for insightful discussions and for a critical reading of an earlier version of the manuscript. PS acknowledges the support of the Anusandhan National Research Foundation (ANRF) under the Science and Engineering Research Board (SERB) Core Research Grant (Grant No.\ CRG/2023/008980). PS also acknowledges support from Vellore Institute of Technology through its Seed Grant (No. SG20230079, Year 2023). GK acknowledges financial support from Vellore Institute of Technology through its Seed Grant (No. SG20230035, 2023).

\appendix

\section{Cancellation of Mixed Terms in the Spatial Contribution}\label{sec:cancel_mix}

In this appendix, we explicitly demonstrate the cancellation of the mixed terms appearing in the spatial part of the Heisenberg equation of motion. for a scalar field in curved spacetime.

We begin with the expression
\begin{align}\label{eq:A1}
- \frac{1}{\sqrt{-g}} \partial_i \left( \frac{g^{0i}}{g^{00}} \hat{\pi} \right)
+ \frac{1}{\sqrt{-g}} \partial_j \left[
\sqrt{-g} \left( \frac{g^{0i} g^{0j}}{g^{00}} - g^{ij} \right)
\partial_i \hat{\phi}
\right].
\end{align}

Substituting the definition of the canonical momentum,
\begin{align}
\hat{\pi} = \sqrt{-g} \left( g^{00} \partial_t \hat{\phi} + g^{0k} \partial_k \hat{\phi} \right),
\end{align}
we obtain
\begin{align}
\frac{g^{0i}}{g^{00}} \hat{\pi}
= \sqrt{-g} \left( g^{0i} \partial_t \hat{\phi}
+ \frac{g^{0i} g^{0k}}{g^{00}} \partial_k \hat{\phi} \right).
\end{align}
Substituting the above into the first term of Eq.(\ref{eq:A1}) gives
\begin{widetext}
\begin{align}\label{eq:A4}
- \frac{1}{\sqrt{-g}} \partial_i \left( \frac{g^{0i}}{g^{00}} \hat{\pi} \right)
&= - \frac{1}{\sqrt{-g}} \partial_i \left[
\sqrt{-g} \, g^{0i} \partial_t \hat{\phi}
+ \sqrt{-g} \, \frac{g^{0i} g^{0k}}{g^{00}} \partial_k \hat{\phi}
\right]\nonumber\\
&= - \frac{1}{\sqrt{-g}} \partial_i \left( \sqrt{-g} \, g^{0i} \partial_t \hat{\phi} \right)
- \frac{1}{\sqrt{-g}} \partial_i \left( \sqrt{-g} \, \frac{g^{0i} g^{0k}}{g^{00}} \partial_k \hat{\phi} \right).
\end{align}

The second term in Eq. (\ref{eq:A1}) becomes
\begin{align}\label{eq:A5}
\frac{1}{\sqrt{-g}} \partial_j \left[
\sqrt{-g} \left( \frac{g^{0i} g^{0j}}{g^{00}} - g^{ij} \right)
\partial_i \hat{\phi}
\right]
= \frac{1}{\sqrt{-g}} \partial_j \left(
\sqrt{-g} \, \frac{g^{0i} g^{0j}}{g^{00}} \partial_i \hat{\phi}
\right)
- \frac{1}{\sqrt{-g}} \partial_j \left(
\sqrt{-g} \, g^{ij} \partial_i \hat{\phi}
\right).
\end{align}
\end{widetext}

Substituting Eqs.(\ref{eq:A4}) and (\ref{eq:A5}) in Eq.(\ref{eq:A1}) and focus on the mixed terms involving
$
\frac{g^{0i} g^{0k}}{g^{00}} \partial_k \hat{\phi}
$ we have
\begin{align}
- \frac{1}{\sqrt{-g}} \partial_i \left( \sqrt{-g} \, \frac{g^{0i} g^{0k}}{g^{00}} \partial_k \hat{\phi} \right)
+ \frac{1}{\sqrt{-g}} \partial_j \left( \sqrt{-g} \, \frac{g^{0i} g^{0j}}{g^{00}} \partial_i \hat{\phi} \right).
\end{align}
Now, relabeling dummy indices in the second term ($j \leftrightarrow i$, $i \leftrightarrow k$), we obtain
\begin{align}
\frac{1}{\sqrt{-g}} \partial_i \left( \sqrt{-g} \, \frac{g^{0i} g^{0k}}{g^{00}} \partial_k \hat{\phi} \right),
\end{align}
which exactly cancels the corresponding term from the first contribution. Hence, the mixed terms vanish identically.

The remaining terms are
\begin{align}
- \frac{1}{\sqrt{-g}} \partial_i \left( \sqrt{-g} \, g^{0i} \partial_t \hat{\phi} \right)
- \frac{1}{\sqrt{-g}} \partial_j \left( \sqrt{-g} \, g^{ij} \partial_i \hat{\phi} \right).
\end{align}

These can be combined into the covariant form
\begin{align}
-\frac{1}{\sqrt{-g}} \partial_j \left( \sqrt{-g} \, g^{j\nu} \partial_\nu \hat{\phi} \right),
\end{align}
where $\nu = 0, i$. This establishes the result used in the main text.

\bibliographystyle{apsrev4-2}
\bibliography{ref}

\end{document}